# The umbra-penumbra area ratio of sunspots during the Maunder Minimum

V.M.S. Carrasco[1,2], J.M. García-Romero[3], J.M. Vaquero[2,4], P.G. Rodríguez[3], P. Foukal[5], M.C. Gallego[1,2], L. Lefèvre[6]

[1] Departamento de Física, Universidad de Extremadura, 06071 Badajoz, Spain [e-mail: vmscarrasco@unex.es]

[2] Instituto Universitario de Investigación del Agua, Cambio Climático y Sostenibilidad (IACYS), Universidad de Extremadura, 06006 Badajoz, Spain

[3] Departamento de Lenguajes y Sistemas Informáticos, Escuela Politécnica, Universidad de Extremadura, 10003 Cáceres, Spain

[4] Departamento de Física, Universidad de Extremadura, 06800 Mérida, Spain

[5] Nahant, MA 01908, USA

[6] Royal Observatory of Belgium, 3 Rue Circulaire, 1180 Brussels, Belgium

**Abstract:** The Maunder Minimum (MM) was a prolonged period of low solar activity that occurred between 1645 and 1715. The true level of solar activity corresponding to this epoch is still a matter of debate. In order to compare solar activity during the MM with that of other epochs, we have evaluated the umbra–penumbra area ratio (U/P hereafter) during the MM. Thus, we have analyzed 196 sunspot drawings including 48 different sunspots observed during the period 1660–1709. The mode value of the ratio obtained from the occurrence frequency distribution lies between 0.15 and 0.25. Furthermore, the median and mean values are equal to 0.24 ± 0.07 and 0.27 ± 0.08 with a sigma clipping, respectively. These results are consistent with recent research using more modern data. Higher U/P values mean faster sunspot decay rates. From our results, the almost absence of sunspots during the Maunder Minimum could not be explained by changes in the U/P since the values of the ratio obtained in this work are similar to values found for other epochs.

## 1. Introduction

Astronomers Gustav Spörer and Edward Maunder pointed out that during the end of the 17th century and the beginning of the 18th century, very few sunspots were seen on the Sun (Maunder, 1890; Spörer, 1887). Eddy (1976) reexamined the records available for that epoch and concluded that this 70-year period (1645 – 1715) was characterized by a prolonged period of low solar activity. This phenomenon is known as the Maunder Minimum (MM) and it is the only grand minimum of solar activity registered by direct telescopic observations.

Other studies published after the benchmark article by Eddy (1976) have provided new information about the MM. Ribes & Nesme-Ribes (1993) built a latitude-time "butterfly" diagram of sunspot occurrence for the MM showing a strong hemispheric asymmetry since sunspots were generally observed in the southern hemisphere. Hoyt & Schatten (1998) recovered thousands of sunspot observations made during the MM, which were included in the construction of the group sunspot number index. Hoyt and Schatten's database has a good temporary coverage for the MM with at least one sunspot record available for more than 95% of days. However, most of these records were based on generic statements on the absence of spots during long periods (Vaquero *et al.*, 2011). It can also be seen in that database that sunspots were rarely observed during the MM without a visible 11-year solar cycle. Moreover, the transition between normal solar activity previous to the MM and low solar activity during the MM was sudden while the recovery from low to normal solar activity after the MM was gradual. More recently, Vaquero *et al*. (2016) carried out a revision of the sunspot group numbers corresponding to the telescopic era, including new information about sunspot records for the MM.

Recent work has provided new knowledge about the MM. Vaquero *et al*. (2015) proposed a 9-year solar cycle during the MM using subsets of the sunspot group database compiled by Hoyt & Schatten (1998). Moreover, Poluianov *et al*. (2014) exhibited a normal 11-year cycle from the reconstruction of solar activity according to cosmogenic isotope data in terrestrial archives. Vaquero *et al*. (2011) concluded, from new sunspot records and a review of some solar observations previously available, that the solar activity level corresponding to the two solar cycles before the beginning of the MM was overestimated. This result would imply a gradual transition from normal to low solar activity previous to the MM. Clette *et al*. (2014) demonstrated that the sunspot group database compiled by Hoyt & Schatten (1998) contains some problems due to several sunspot records that were obtained from solar meridian observations. Vaquero & Gallego (2014), after analyzing a set of solar meridian observations (accompanied with notes about sunspots made at the Royal Observatory of the Spanish Navy, San Fernando, Spain) from the period 1833–1840, showed the solar activity obtained from this kind of record could be underestimated.

Comprehension of the characteristics of a grand minimum period such as the MM has great importance in several scientific fields to better understand long-term solar activity and its influence on the heliosphere or on Earth. Some recent analysis from cosmogenic isotope data proposed that the MM conditions could return within 50 years (Lockwood, 2010). The level of solar activity during the MM is still a matter of debate. Hoyt & Schatten (1998) obtained a low solar activity level in their reconstruction of the group sunspot index. However, in recent work (Rek, 2013; Zolotova & Ponyavin, 2015) it has been proposed that the MM could be not a grand minimum period of solar activity but a secular minimum. For example, Zolotova & Ponyavin (2015) argued that the solar activity level during the MM could be underestimated because objects with irregular shapes on the solar surface could have been deliberately omitted in the textual reports. Zolotova & Ponyavin (2015) suggested that such omission can be caused by the dominant worldview of the 17$^{th}$ century that sunspots (Sun's planets) are shadows from transits of unknown celestial bodies. However, Usoskin *et al*. (2015) identified serious errors in the study of Zolotova & Ponyavin (2015) and concluded that solar activity level was exceptionally low during the MM, rejecting the ideas of a moderate or high level of solar activity during the MM. Furthermore, Carrasco *et al*. (2015) and Carrasco & Vaquero (2016) obtained low values for the solar activity level compatible with a grand minimum of solar activity from sunspot observations made by Hevelius and Flamsteed during the MM.

Hoyt & Schatten (1997) noted that the rate of sunspot decay is proportional to the convective velocity. They pointed out that higher convective velocities lead to higher values for the U/P and faster sunspot decay rates. Therefore, according to those statements, the ratio between the areas of the umbra and penumbra of sunspots is an indicator of sunspot decay rate and indirectly related to the solar irradiance variation (Hoyt & Schatten, 1997). In one of the first studies of the U/P, Nicholson (1933) estimated its value as 0.21 using data from the Royal Greenwich Observatory. Waldmeier (1939) analysed some sunspot photographs taken by Wolfer obtaining a similar ratio but with values from 0.15 to 0.3 (increasing with the sunspot size). Recently, Vaquero *et al*. (2005) found an average ratio of 0.255 from the de la Rue data belonging to the period 1862–1866. Hathaway (2013) showed that the penumbra–umbra ratio (inverse to the studied here U/P) increases from 5 to 6 as the sunspot group area increases from 100 to 2000 millionths of solar hemisphere from Royal Greenwich Observatory records (1874–1976). Moreover, he showed that this ratio does not vary with the sunspot group latitude or the phase of the solar cycle but found a 100-year secular variation in the penumbra-umbra ratio for sunspot groups with area values lower than 100 millionths of solar hemisphere). Hathaway (2013) showed that the ratio values decreased smoothly from 7 in 1905 to lower than 3 around 1930 and then smoothly increased to 7 until 1961. However, the secular variation in the penumbra-umbra ratio for small sunspot groups was not confirmed by Carrasco *et al*. (2018) using the Coimbra sunspot catalogue.

The aim of this manuscript is to evaluate the U/P during the MM to check whether there are changes in this parameter with respect to that during other epochs. Thus, 196 drawings corresponding to 48 different sunspot groups performed during the MM have been analyzed. We present these observations in Section 2. An explanation of the methodology can be found in Section 3. Section 4 is devoted to showing the results, and the main conclusions are discussed in Section 5.

## 2. Data

We performed an automatic analysis of 196 sunspot drawings to determine the relationship between penumbral and umbral area of 48 sunspot groups observed during the MM. These sunspot records were obtained during the period 1660–1709 by 13 different observers, including one, whose name is presently unknown. The Parisian astronomers P. de la Hire and G.D. Cassini are the observers with the largest number of sunspot records studied in this work. We examined 70 sunspot records (13 different sunspot groups) spread over 9 historical sources according to the de la Hire observations and 46 (11 different sunspot groups) in 4 historical sources with respect to the de la Hire records. For each source, one author is generally assigned as the only observer responsible of the records included in that document, but there are five cases where two authors are appointed. All the observers are responsible for at least some observation individually, except for Fogelius who shares all of his observations with another observer, G.D. Cassini.

The images examined were extracted from 24 documents listed in Table 1, which provides for each documentary source: i) the code that we assigned to that source, ii) the author of the sunspot drawings, iii) the observation period of the sunspots, and iv) the number of sunspots examined for that documentary source and author. We note that in the last case in Table 1, there are sunspots observed by G.D. Cassini in 1676 and other sunspot observations made by an unknown observer in 1684. The codes assigned to each documentary source follow the format XX_YY_ZZ, where XX is always MM (Maunder Minimum), YY the initials of the name of each documentary source, and ZZ is composed by numbers starting from 01 to distinguish the different documents analyzed for the same source. Moreover, we assigned different numbers, starting from 01, to the distinct sunspots registered in the same document and another additional index, starting from 01, to the different records carried out for the same sunspot. For example, the code assigned to the second sunspot registered by G.D. Cassini on 27 November 1676 included in the historical source *Memoires de l'Academie Royale des Sciences* is MM_AS_01_02_09. A complete report on all sunspots analysed in this work including information on date, observer, source and the ratio value is publicly available at the website of the Historical Archive of Sunspot Observations (http://haso.unex.es/). An example of sunspots examined in this work is shown in Figure 1.

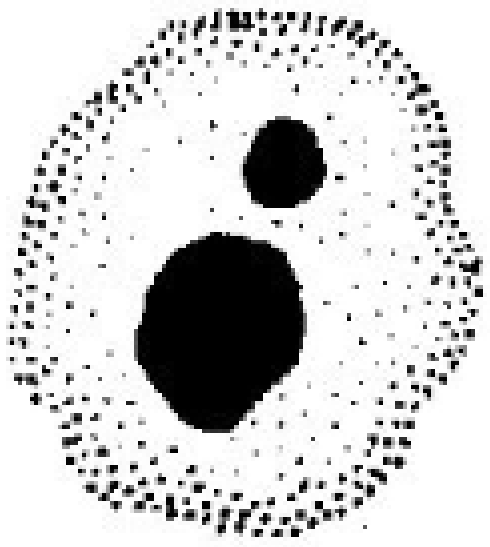

Figure 1. An example of the sunspots examined in this work. This sunspot was recorded by P. de la Hire on June 22nd 1703 [Source: *Memoires de l'Academie Royale des Sciences*]. The code assigned in this work to this sunspot is MM_AS_08_02_05.

Table 1. Historical documentary sources used to compute the U/P during the MM. The code assigned for this study includes the author responsible of the sunspot drawings, observation period and number of sunspots for each documentary source.

| SOURCE | CODE | AUTHOR | PERIOD | SUNSPOTS |
|---|---|---|---|---|
| Acta Eruditorum | MM_AE_01 | Jartoux | 1701 | 12 |
| Acta Eruditorum | MM_AE_02 | Feuillée | 1709 | 1 |
| Journal des Observations | MM_JO_01 | Feuillée | 1709 | 4 |
| Journal des Sçavans | MM_JS_01 | G.D. Cassini | 1676 | 19 |
| Memoires de l'Academie Royale des Sciences | MM_AS_01 | G.D. Cassini | 1676 | 19 |
| Memoires de l'Academie Royale des Sciences | MM_AS_02 | G.D. Cassini | 1678 | 3 |
| Memoires de l'Academie Royale des Sciences | MM_AS_03 | G.D. Cassini | 1684 | 4 |
| Memoires de l'Academie Royale | MM_AS_04 | G.D. Cassini | 1701 | 1 |
| des Sciences | | Maraldi | | |
| Memoires de l’Academie Royale des Sciences | MM_AS_05 | Cassini (son) | 1700-1701 | 5 |
| Memoires de l’Academie Royale des Sciences | MM_AS_06 | Cassini (son) De la Hire | 1702 | 4 |
| Memoires de l’Academie Royale des Sciences | MM_AS_07 | Cassini (son) | 1703 | 5 |
| Memoires de l’Academie Royale des Sciences | MM_AS_08 | De la Hire | 1703 | 18 |
| Memoires de l’Academie Royale des Sciences | MM_AS_09 | De la Hire | 1703 | 19 |
| Memoires de l’Academie Royale des Sciences | MM_AS_10 | Maraldi | 1704 | 6 |
| Memoires de l’Academie Royale des Sciences | MM_AS_11 | De la Hire Maraldi | 1704 | 5 |
| Philosophical Transactions | MM_PT_01 | G.D. Cassini | 1671 | 4 |
| Philosophical Transactions | MM_PT_02 | Hevelius | 1671 | 9 |
| Philosophical Transactions | MM_PT_03 | G.D. Cassini | 1676 | 6 |
| Philosophical Transactions | MM_PT_04 | Stannyan | 1703 | 9 |
| Philosophical Transactions | MM_PT_05 | G.D. Cassini Fogelius | 1671 | 8 |
| Philosophical Transactions | MM_PT_06 | Hook | 1660-1671 | 2 |
| Bion | MM_BI_01 | Bion | 1672 | 5 |
| Le Monnier | MM_LM_01 | Picard | 1676 | 13 |
| Le Monnier | MM_LM_02 | G.D. Cassini<br>Unknown | 1676<br>1684 | 6<br>9 |

## 3. Method

A software tool in OpenCV format has been developed in order to obtain the U/P from observations made during the Maunder Minimum (Figure 2). Several problems were encountered because some images were very small, of poor quality, noisy and/or the sheets presented unbalanced whites. However, we have been able to extract, using different combinations of algorithms and parameters, the significant contours of umbra and penumbra.

The images were processed with different algorithms for smoothing, thresholding, morphological, filling, size variation, sampling, etc. For example, the smoothing allows to reduce the noise, eliminate imperfections of the image and outline contours of the umbra and penumbra. Thresholding binarizes the image, that is, represents it with two colors, for example, black and white. Morphological algorithms eliminate imperfections and noise of the images taking into account the shape and structure of the image. Sampling of images is a technique that provides help when the initial characteristics of the image are complicated due, for example, to very few or too many pixels in the image.

The main functions are: i) to upload images in png, jpg or bmp format, ii) to select among several kind of image filters and set the values of their parameters, iii) to execute an automatic application of consecutive filters, iv) to display of execution of the filters to check how each filter affects the input image, v) to change the image display to see the processed image or the significant contours of that image, vi) to display the contour information as area and perimeter, vii) to select maximum and minimum thresholds for the area of the contours in order to search appropriate contours, viii) to check the original image and final result with the contours found after executing the filters, xix) umbra and penumbra execution mode to select two sequences of different filters (one to obtain the umbra and the other for the penumbra), x) to save and load the filter sequences at any time, xi) to generate reports with the final information, xii) export the results to a different formats such as Word, PDF or Excel, and xiii) to set different parameters in the application as metadata relating to the sunspot from “user settings”.

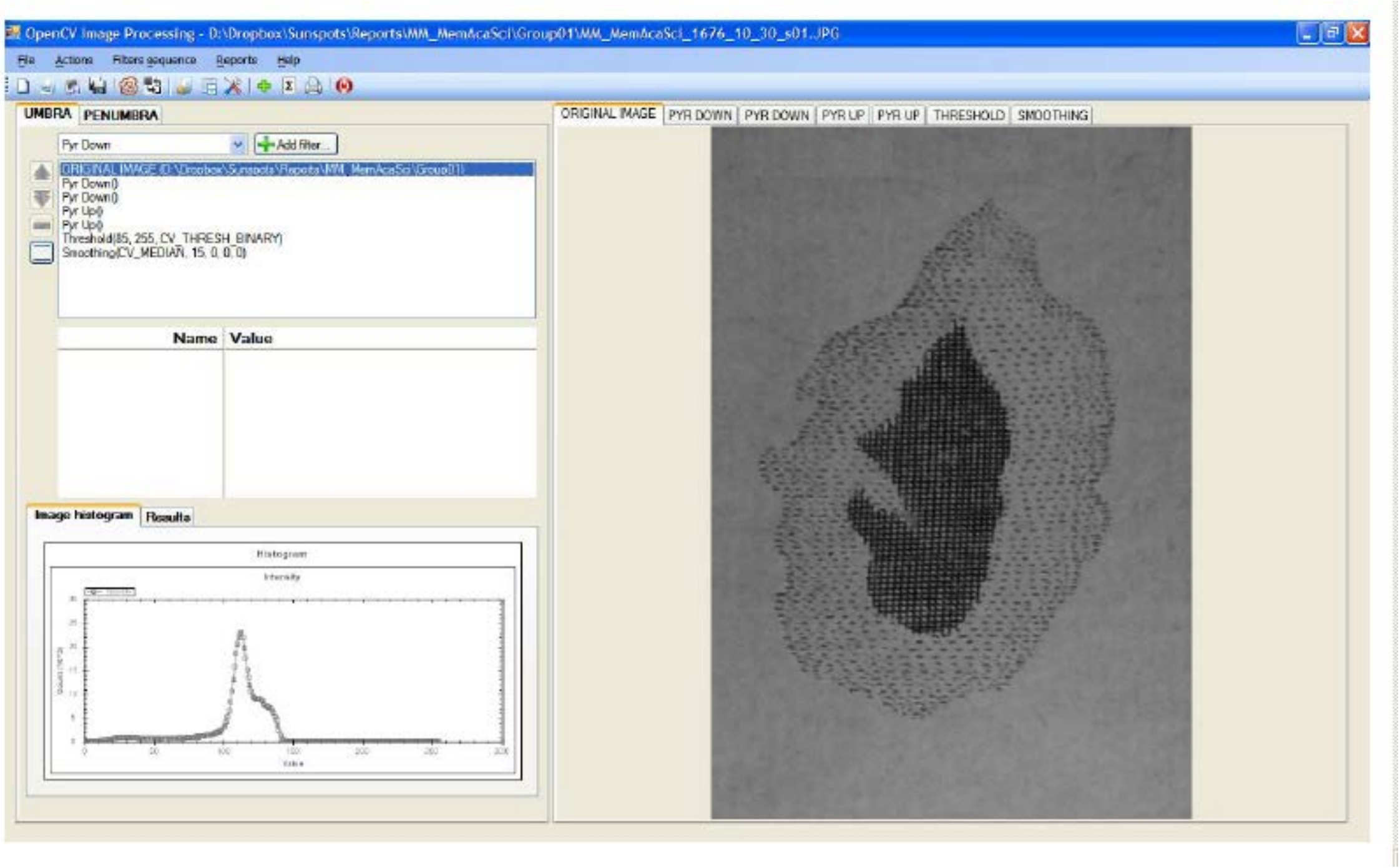

Figure 2. The main screen of the application created in order to obtain the U/P. The code assigned to this sunspot is MM_AS_01_01_01.

## 4. Results and Discussion

### 4.1. Discussion on the U/P during the Maunder Minimum and its comparison with Debrecen Heliophysical Observatory data

In this work, we examined 196 sunspot drawings by 13 different authors in order to compute the U/P during the MM. Figure 3 shows the U/P ratio for all sunspot drawings analyzed. We note that, in this analysis, the following ratio values were not included: i) the sunspot observations made by G.D. Cassini from 1676 October 30 to 1676

November 30 included in the *Memoires de l'Academie Royale des Science* and Le Monnier (1741) since *Journal des Sçavans* contains the same observations and the definition of the sunspot images in this documentary source is better, and ii) the sunspot observation made by Feuillée on 1709 January 30 registered in *Journal des Observations* because it is also included in *Acta Eruditorum* with better image quality.

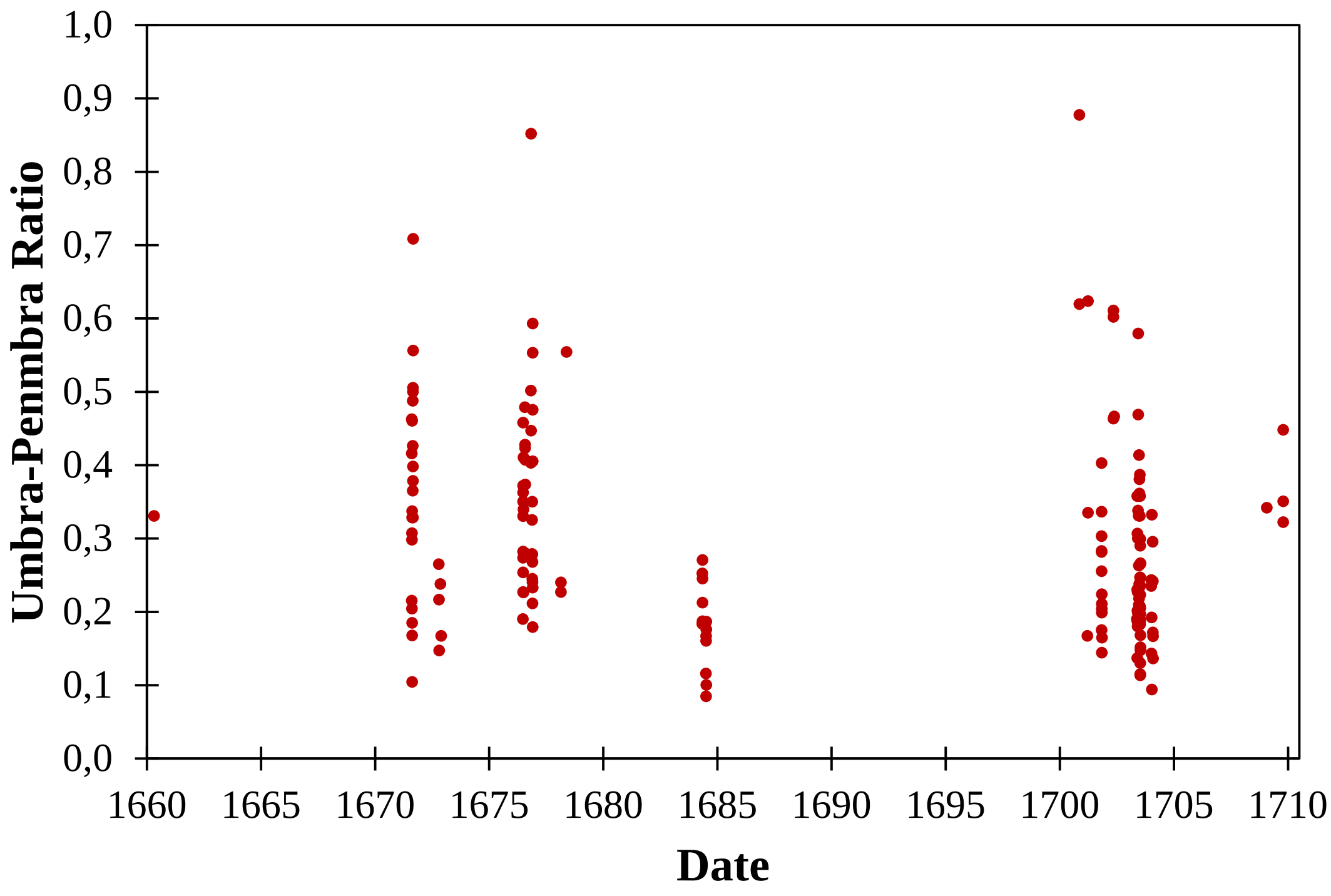

Figure 3. Evolution of the U/P for all sunspots analyzed in this work.

In order to compute the mean U/P for all sunspots analyzed in this work, we represent all the umbral areas *versus* penumbral areas (Figure 4). The best linear fit for these data is given by: $A_u = (0.25 \pm 0.01) A_p + (1032 \pm 725)$, $r = 0.87$, $p$-value $< 0.001$. Unfortunately, we do not know in most cases the relationship between pixel size and area in millionths of hemisphere or disk. Therefore, we can only plot Figure 4 using pixels, which depend on the characteristics of each image. A better estimate of U/P for all sunspots analyzed in this work is shown in Figure 5 (left panel) where a histogram is presented using bins of 0.05 units. The mode of the distribution lies for the ratios equal to 0.15-0.2 and 0.2-0.25. These two bins have the same number of occurrences (33).

These values agree with ratio values obtained in other works for other epochs (Vaquero *et al*., 2005).

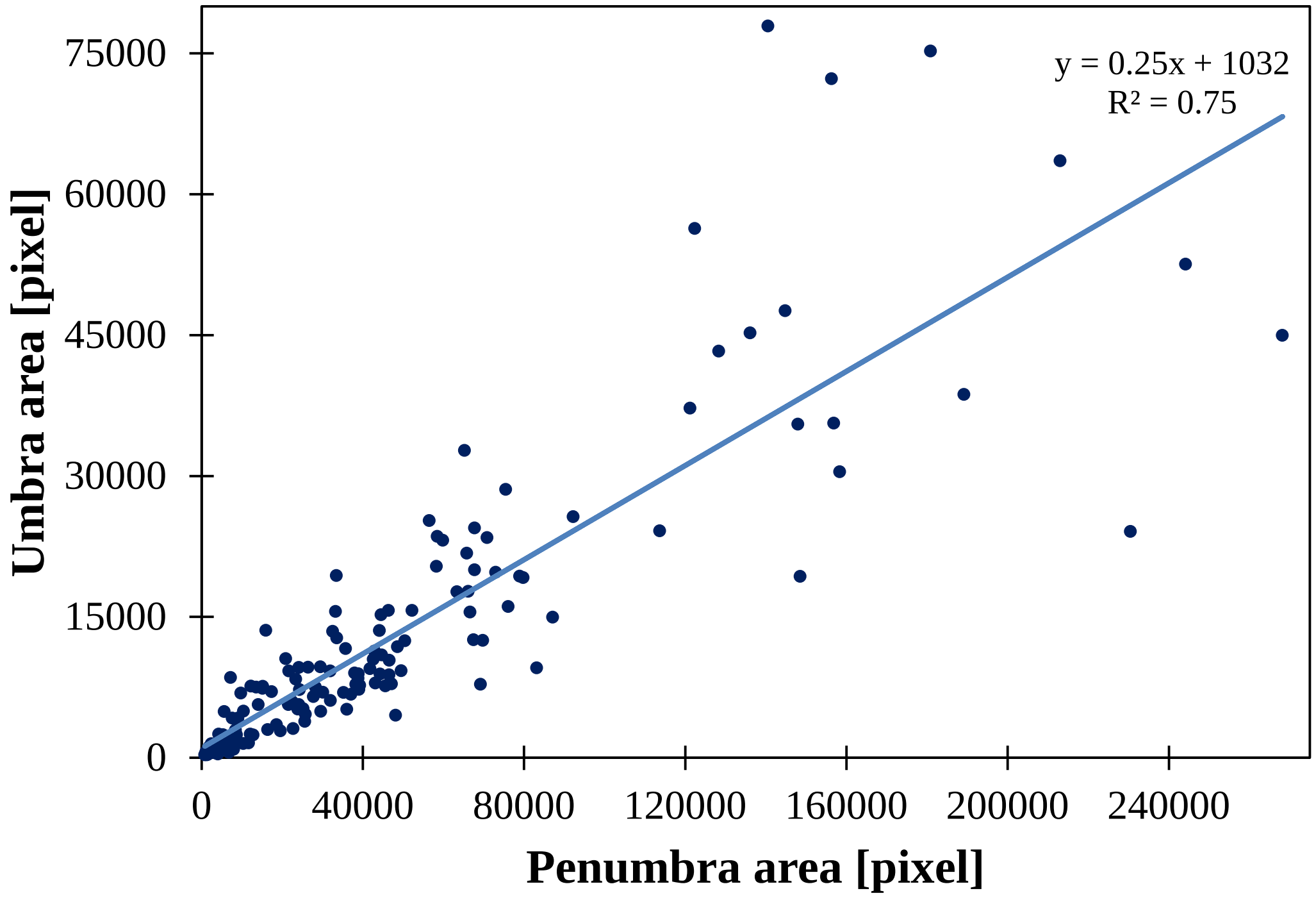

Figure 4. Linear relationship between umbral and penumbral area. The units of these areas are given by pixels and the best linear fit equation and R-squared coefficient are shown.

In addition, we compared our results with the U/P computed from individual sunspots recorded in the Debrecen Heliophysical Observatory catalog (http://fenyi.solarobs.csfk.mta.hu/DPD/index.html) (hereafter DPD) for the period 1982-2010. In order to compare both distributions, we calculated the median values and standard deviations of the U/P, applying a sigma clipping to avoid the effect of the outliers. We stopped the iteration of the sigma clipping when the new measured standard deviation ($\sigma_{new}$) is within of a tolerance level of the old one ($\sigma_{old}$) defined by: $(\sigma_{old} - \sigma_{new})/\sigma_{new} < 1$. We chose the median and not the mean to perform the sigma clipping because the median is less affected by the presence of outliers. Discarding in this way the outliers from the distributions, we obtained 0.18 ± 0.06 (0.22 ± 0.02) in the case of DPD and 0.24 ± 0.07 (0.27 ± 0.08) for the MM distribution according to median (mean) values. These values are compatible within the error bars both median and mean values and, moreover, they agree with U/P obtained for other epochs (Vaquero *et al*., 2005). On the other hand, the mode of the distribution using Debrecen data lie for ratio values between 0.15 and 0.2 while it is between 0.15 and 0.25 according to this work from sunspot observation during the MM. Figure 5 also shows a difference between both distributions. The distribution according to Debrecen data presents a greater frequency for low U/P values than the distribution obtained from MM records. Furthermore, we apply the Student's t-test with a 95 % confidence interval in order to study the similarity of the distributions. The results of this statistical test is $t = -7.36$, $p$-value < 0.001. According to the result, the absolute value of the calculated $t$-value is greater than the critical $t$-value (1.98) at 95 % confidence interval. Therefore, there is significant statistical difference between both distributions. Nevertheless, if we discard the U/P values lower than 0.1 in both distributions and consider a 99.9 % confidence interval, the obtained $t$-value (3.94) becomes close to the critical $t$-value of 3.36. This implies that the difference between both distributions can be largely explained by the natural observational bias towards larger sunspots observed in the MM observations. Moreover, there are other factors that can contribute to the difference: i) low number of statistics in the MM distribution in comparison with the number of data in the DPD distribution; ii) differences in the instruments employed to observe (DPD has modern telescopes with a better resolution); iii) the observations in DPD were carried out by a large number of observers, even with observations from other observatories, with respect to the 13 observers for MM distribution. We want to highlight that in spite of the statistical difference between both distributions, the mode, median, and mean values are similar.

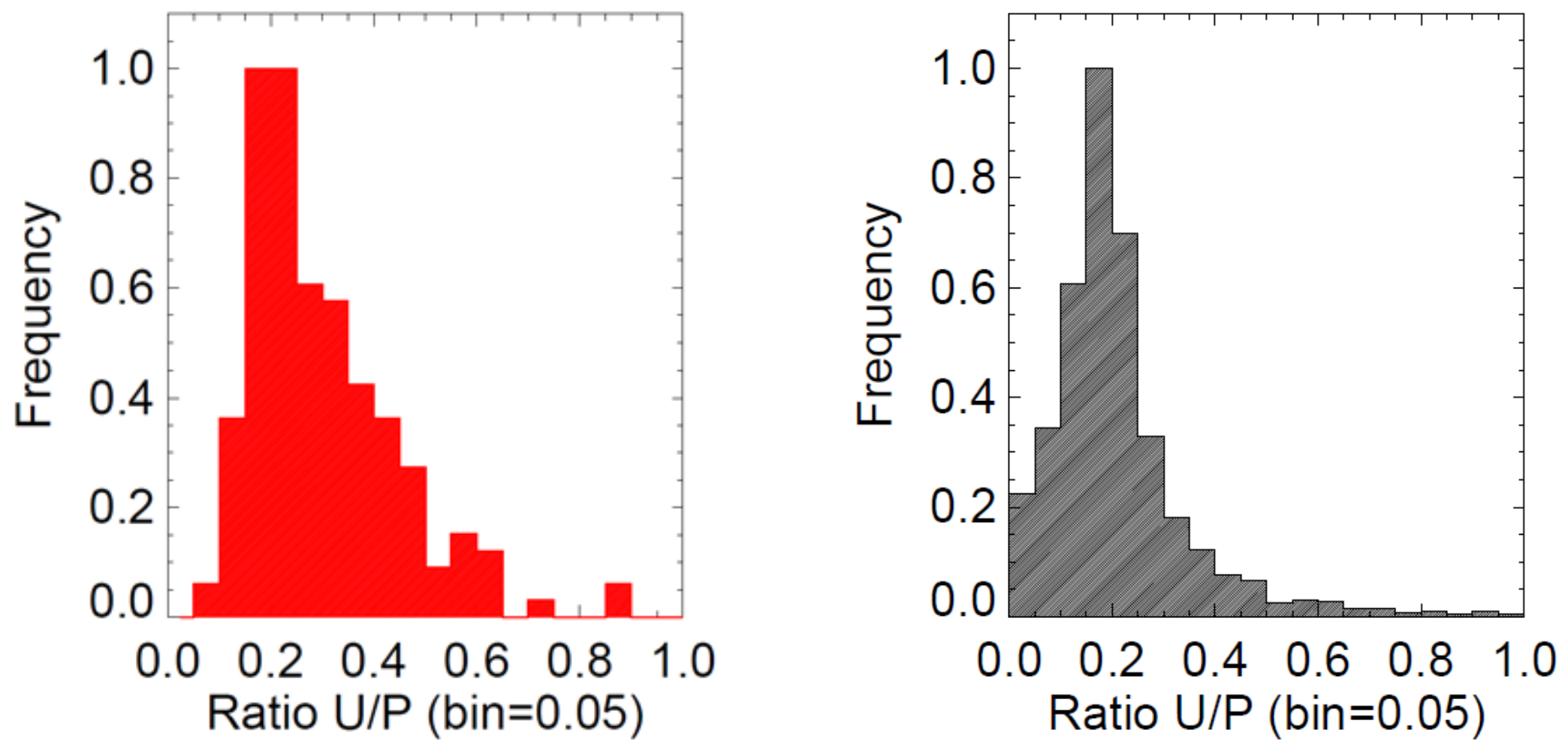

Figure 5. Comparison between normalized histograms for the values of the U/P from sunspot groups analysed in this work (left panel) and those registered by the Debrecen Heliophysical Observatory (DPD) for the period 1982-2010 (right panel). Ratio data are represented in bins of 0.05 units.

### 4.2. Comparison between different observers during the Maunder Minimum

The average U/P for all the days in which the same sunspot was registered by one or several observers is depicted in Figure 6. The different colors indicate the observers responsible for those observations. Five of the documents examined in this work have two observers as responsible of the same observations (see Table 1) and therefore, in Figure 6, the values of the U/P are shown in the series of both authors.

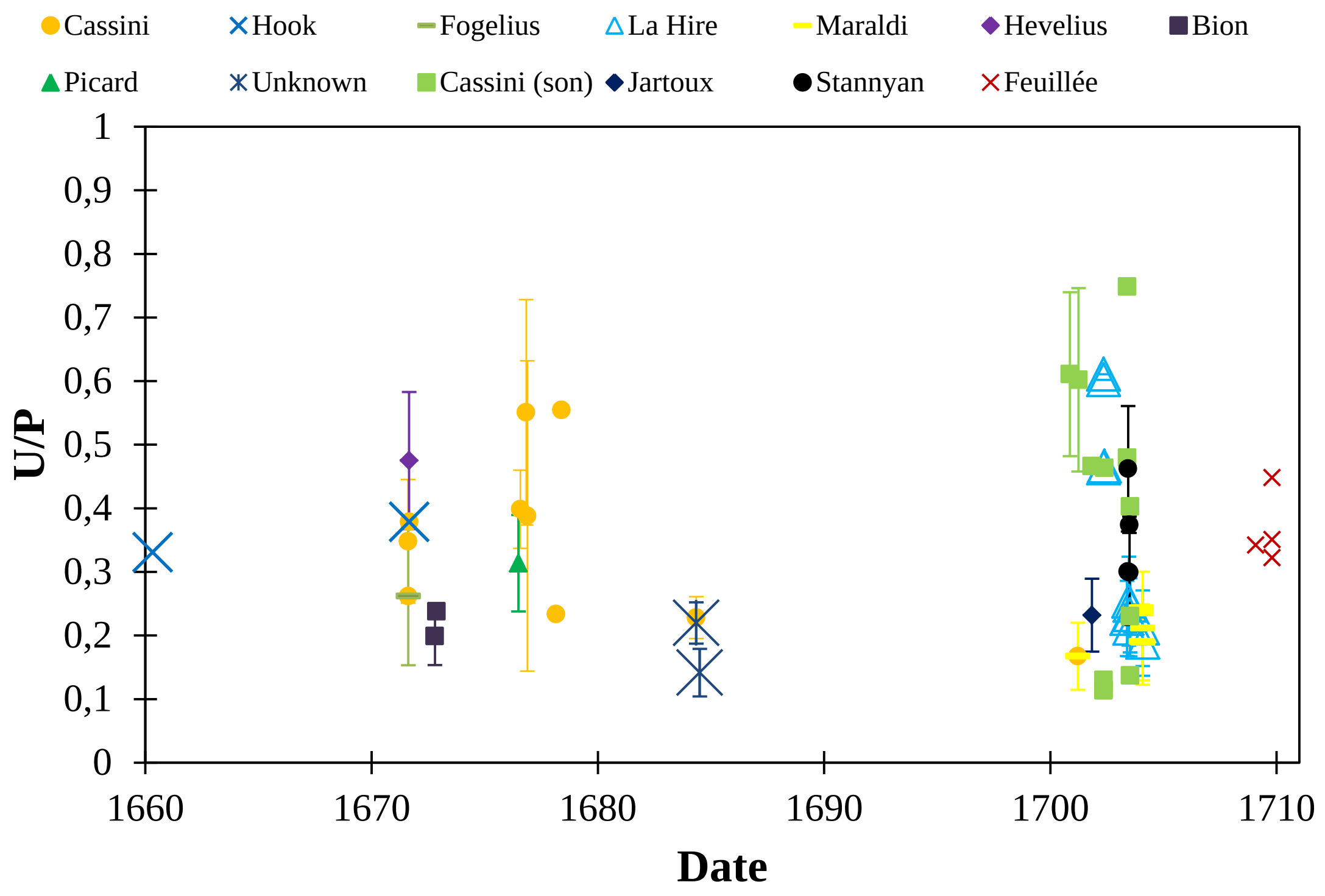

Figure 6. Average U/P of each sunspot studied in this work recorded by the 13 different observers. Error bars represent one standard deviation.

We can study the evolution of the U/P as the sunspot crosses the solar disk. Thus, we selected the cases where the same sunspot was observed during at least eight different days (Figure 7). The most striking example was observed by G.D. Cassini from 1676 November 18 to 1676 November 30 (red line in Figure 7). In this case, we can see evolution of U/P from its appearance at the eastern limb to its disappearance in the west. We can see that the value of the ratio when the sunspot is close to the solar limbs sharply increases due to foreshortening effect. In the rest of the trajectory, in general, the ratio presents values around 0.25. Similar behavior seems to occur for a spot observed by Hevelius in 1671 (purple line) and for another observed by de la Hire in 1703 June (green line). Considering the sunspot observed by de la Hire between 1703 May 20 and 1703 June 3 (black line), the foreshortening effect can be seen in the first observation day when the sunspot is next to the eastern limb but not when the sunspot is close to the opposite limb. We note that there is a difference of 5 days between the de la Hire's first and second observation. In his other two records, the ratio values are more stable although in the last drawing of each sunspot, corresponding on 1703 July 16 and 15 for the blue and orange line respectively, those sunspots were close to the western limb of the Sun.

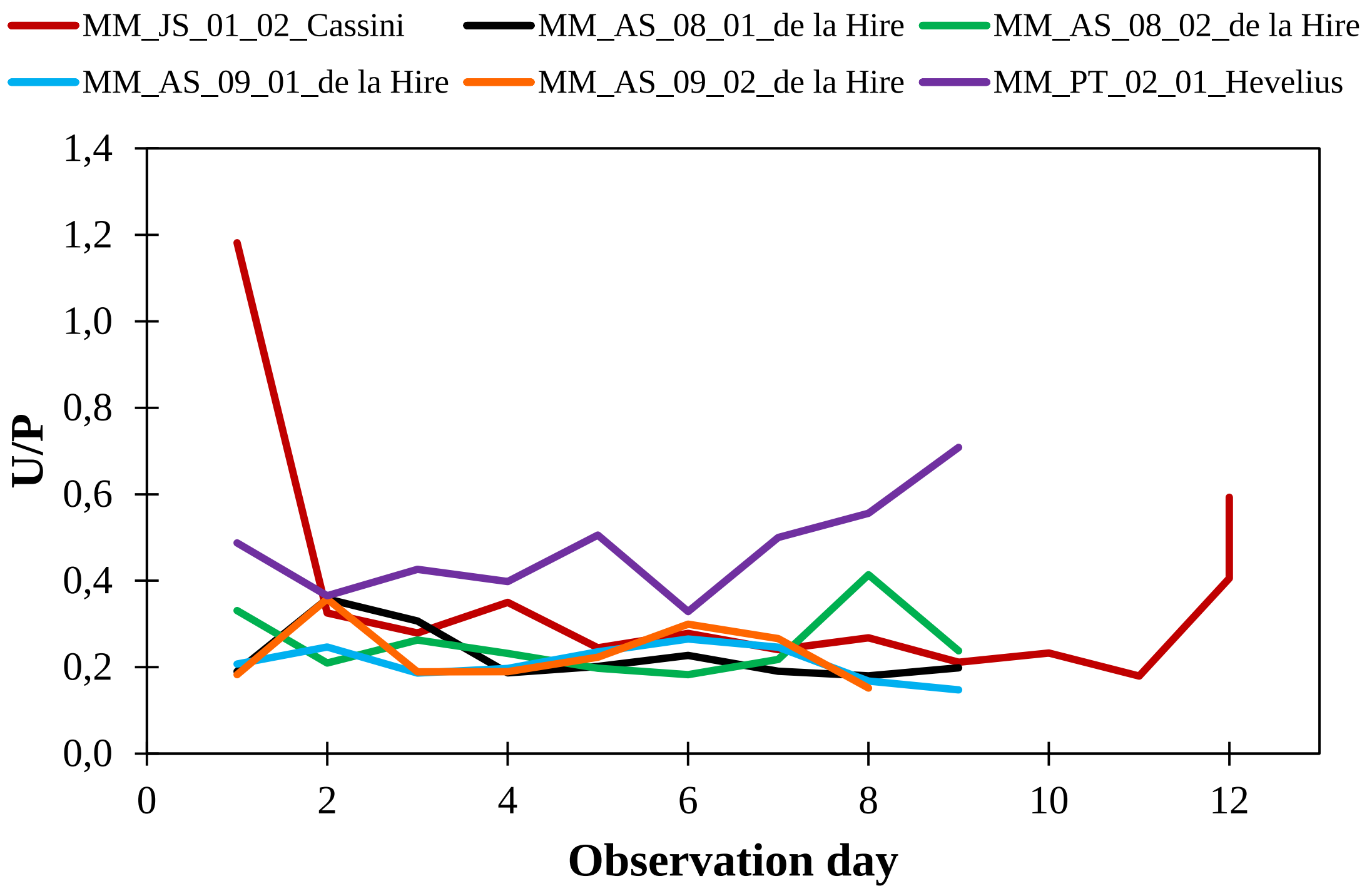

Figure 7. Evolution of the U/P of a sunspot crossing the solar disk. Colored lines represent U/P from the first observation day of the sunspot for the cases of: G.D. Cassini (1676 November 18 – red line), de la Hire (1703 May 20 – black line, 1703 June 18 – green line, 1703 July 8 – blue line, 1703 July 8 – orange line), and Hevelius (1671 August 26 – purple line). On top, codes assigned to different historical documentary sources for this work are shown (see Table 1).

In order to compare U/P values obtained from sunspot drawings made by several observers, we selected two cases where different observers registered the same sunspot (Figure 8). The first case is that of the sunspot observed by G.D. Cassini and an unknown observer in 1684 May (Figure 8, top panel). The average U/P according to that observation dataset made by G.D. Cassini is equal to 0.23±0.03 while it is 0.22±0.03 in the case of unknown observer. When both observers have one record the same day (1684 May 5), the value of the ratio obtained by the unknown observer is slightly greater than the value obtained by G.D. Cassini. The average value of the ratio using observations by G.D. Cassini is 25% smaller. The other case concerns three different sunspots registered by Cassini (son) and de la Hire in 1703 July (Figure 8, bottom panel). We only have one observation per day for these three sunspots in the Cassini (son) data, on 1703 July 7, while we have observations of these sunspots for different days from the de la Hire records. The ratio values computed from Cassini (son) for these three sunspots on 1703 July 7 are 0.12, 0.13, and 0.11, a 40-55 % lower than those obtained from the de la Hire data for the same day: 0.19, 0.29, and 0.20, respectively. We note that the average ratio values for all observations days of these three sunspots according to the de la Hire drawings are equal to 0.23±0.07, 0.25±0.04, and 0.21±0.04, respectively. The sizes of the umbrae drawn by la Hire are significantly bigger than those drawn by Cassini (son) although Cassini (son) registered a higher number of umbrae for the same sunspot.

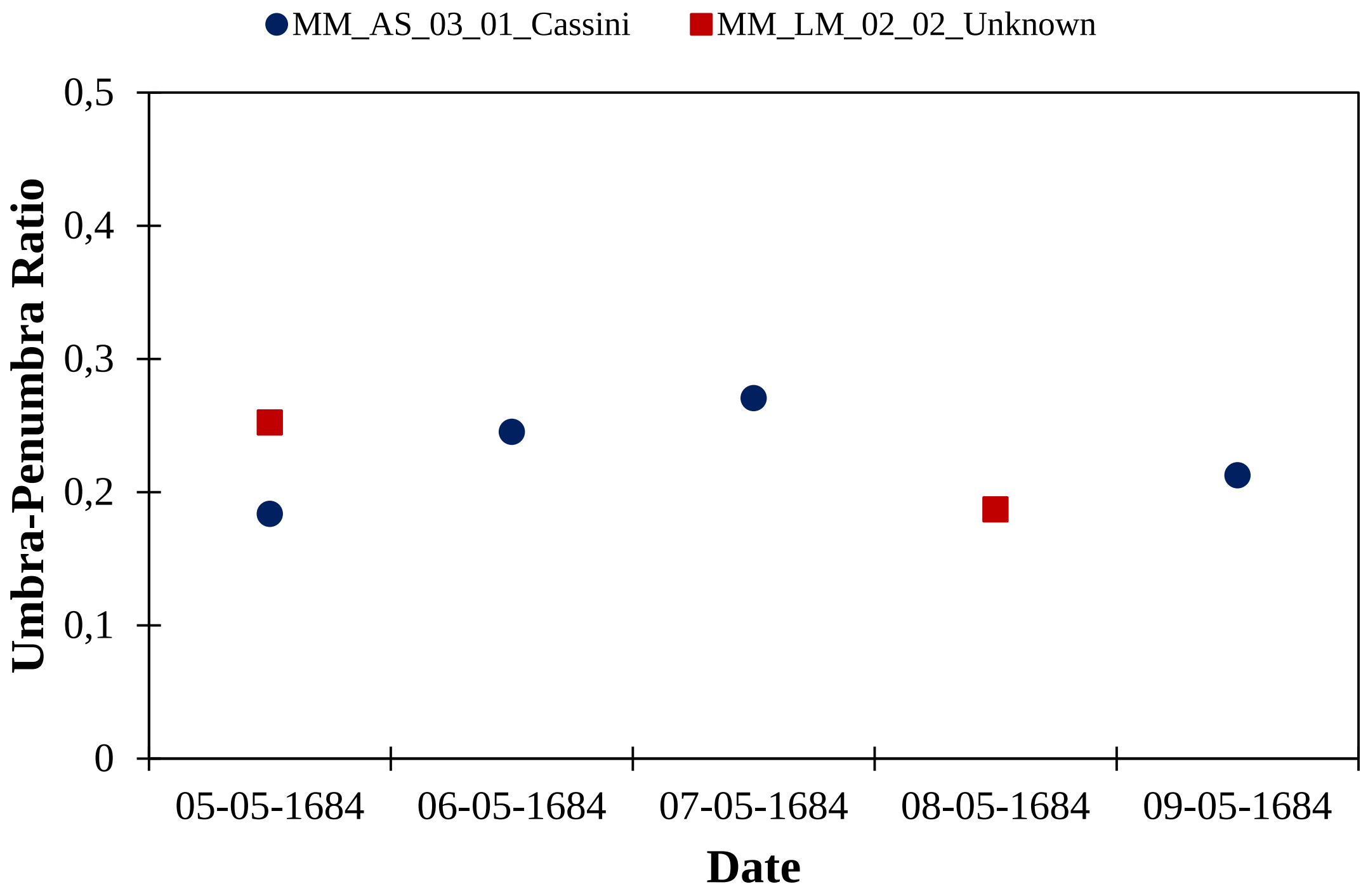

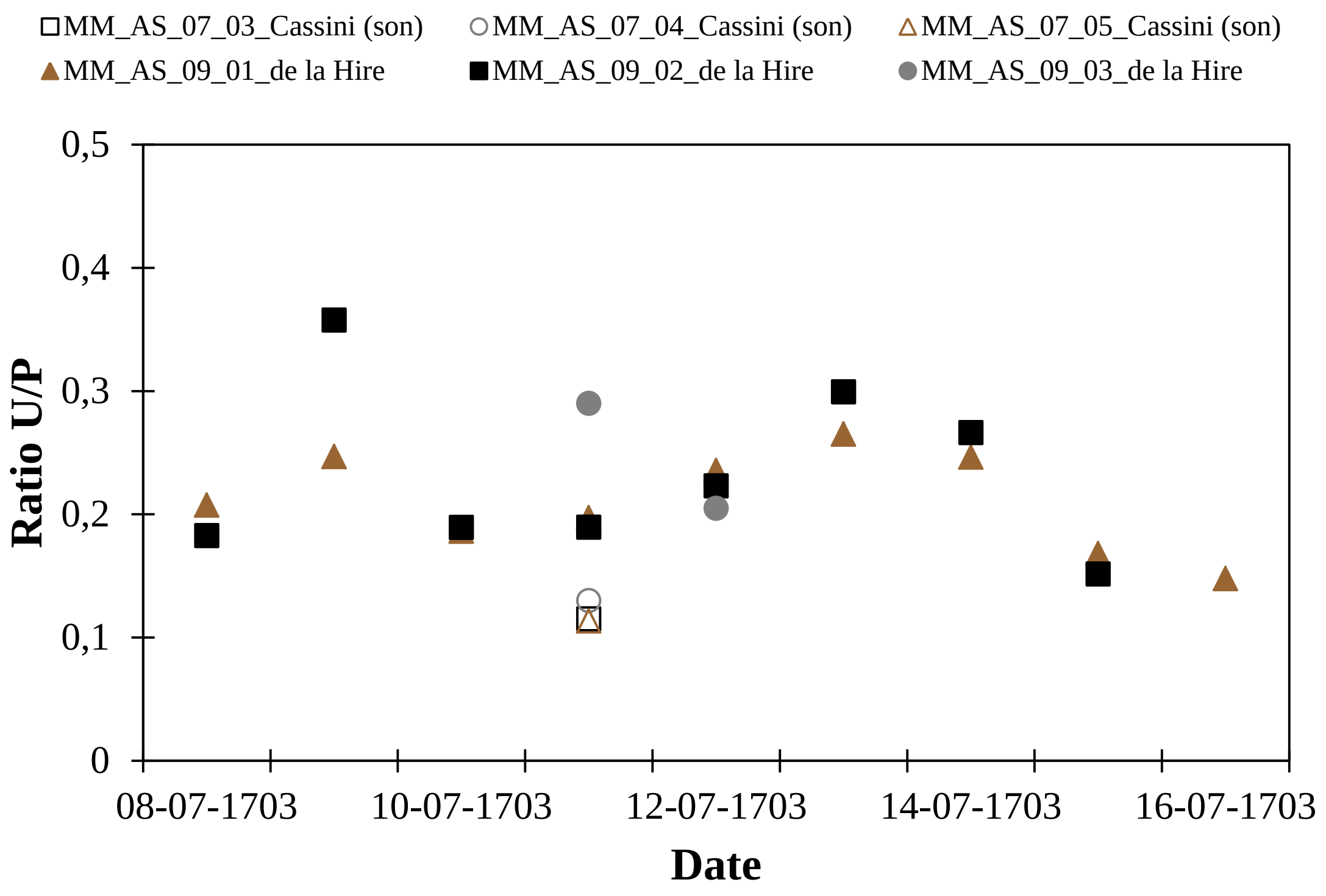

Figure 8. Comparison of the U/P computed from sunspot drawings made by: i) Top panel - G.D. Cassini (blue circles) and unknown observer (red squares) on 1684 May, and ii) Bottom panel - Cassini (son) (empty symbols) and de la Hire (full symbols) in 1703 July. In the last case, each symbol (full or empty squares, circles, and triangles) represents the evolution of the same sunspot. Codes assigned to different historical documentary sources for this work are shown on the top (see Table 1).

We found sunspots observed by G.D. Cassini from 1676 October 30 to 1676 November 30 in three different historical documentary sources: *Journal des Sçavans*, *Memoires de l'Academie Royale des Sciences*, and one more in Bion (1751). We can evaluate the methodology employed in this work comparing the U/P computed from these three different sources. Figure 9 represents the values of U/P calculated according to the sunspot drawings made by G.D. Cassini contained in those historical sources. We note that only 6 records are available from Bion (1751) while 19 records are available for the two other sources. Moreover, the last four ratio values obtained from *Journal des Sçavans* and *Memoires de l'Academie Royale des Sciences*, and the last ratio value computed from Bion (1751) in Figure 9 correspond to the same day of November 30.

We want to highlight that the values obtained from the three documentary sources are very similar. Moreover, the foreshortening effect can be seen in the ratio values both in the last record of the first sunspot (November 3) when value is around 0.9 and at the beginning (November 18) and the end (November 30) of the second period. For the first sunspot, the values of the U/P are high (between 0.35 and 0.5 from data of the three documentary sources but without considering the last day of the period, November 3) probably due to the proximity of the sunspot with respect to the solar limb. For the second period, except for high values due to foreshortening effect (November 18 and 30), the values for this U/P lie between 0.2 and 0.35, approximately. The average value of the ratio for the first sunspot (October 30 – November 3, both included) is equal to 0.55 ± 0.18 and 0.55 ± 0.22 according to records included in *Journal des Sçavans* and *Memoires de l'Academie Royale des Sciences*, respectively. In the case of the second sunspot (November 18 – November 30), the average ratio is equal to 0.39 ± 0.24 and 0.50 ± 0.47 for the documentary sources previously mentioned, respectively. The difference in these values is due to the foreshortening effect. If the values affected by foreshortening are removed, the new ratios are 0.29 ± 0.08 and 0.31 ± 0.10, respectively. Although there are few sunspot records included in Bion (1751) to be compared with the other two, the values of the U/P are similar to those corresponding to the other two sources (Figure 9). In this case, the average U/P is equal to 0.42 ± 0.06 for the first observation period and 0.28 ± 0.07 for the second one.

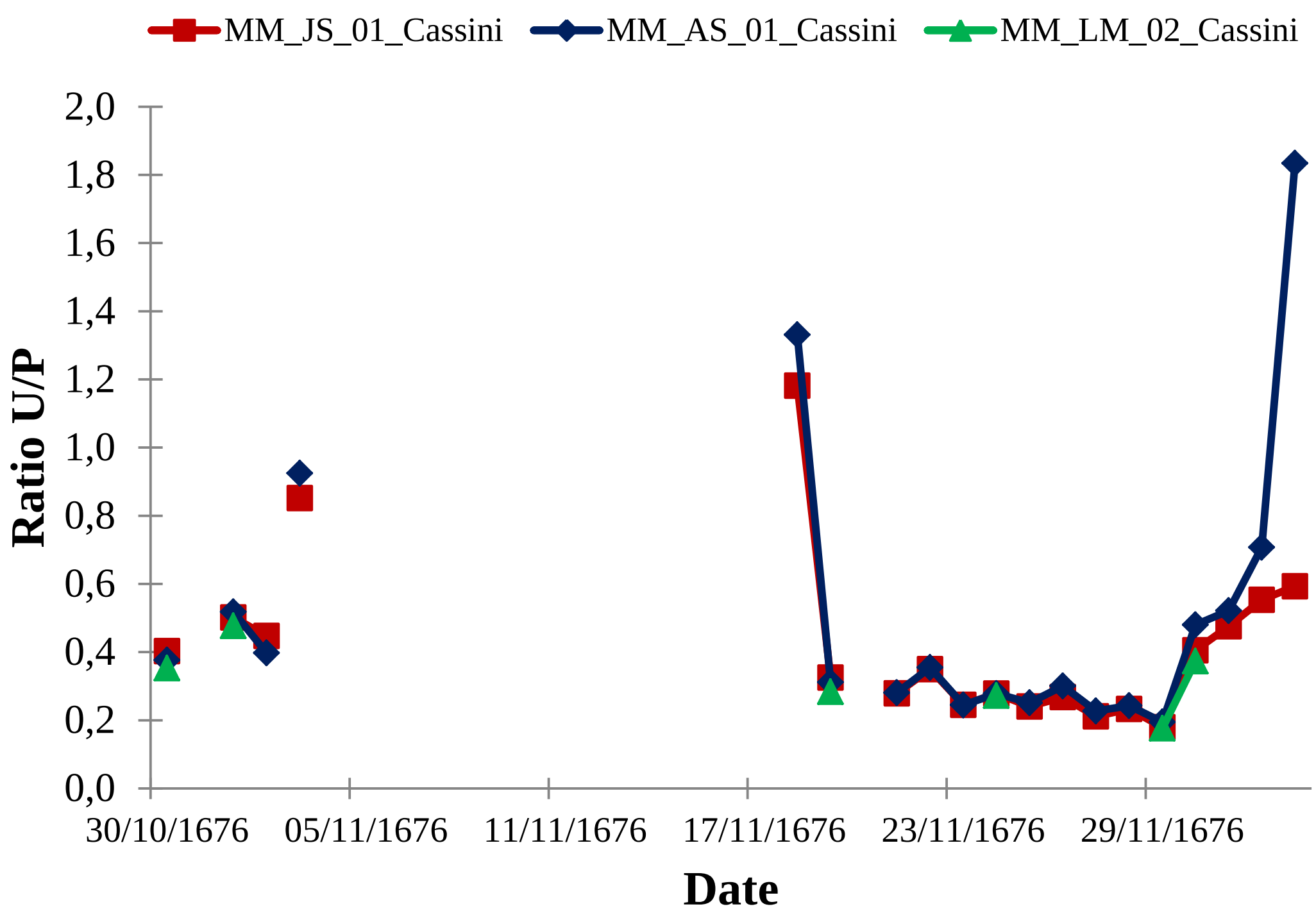

Figure 9. Comparison of U/P computed for the same sunspot registered in three different historical sources. Red line represents the values obtained from *Journal des Sçavans*, blue line from *Memoires de l'Academie Royale des Sciences*, and green line from Bion (1751). Codes assigned to different historical documentary sources for this work are shown on the top (see Table 1). The four last ratio values for blue and red lines and the last ratio value for the green line correspond to November 30.

## 5. Conclusions

We computed the U/P for 196 sunspot drawings that contain 48 different sunspots observed for the period 1660–1709 including the core of the MM. The sunspot observations analysed in this work were made by 13 different observers and the drawings were extracted from 24 documents included in seven historical documentary sources: *Acta Eruditorum*, *Journal des Observations*, *Journal des Sçavans*, *Memoires de l'Academie Royale des Sciences*, *Philosophical Transactions*, Bion (1751), and Le Monnier (1741).

The modes of the occurrence frequency distribution of the U/P values obtained in this work lie for the ratios equal to 0.15-0.25. We calculated the median value and standard deviation of the U/P applying a sigma clipping in order to avoid the effect of the outliers. Then, discarding the outliers, we obtained 0.24 ± 0.07 and 0.27 ± 0.08 for the median and mean values, respectively. We have obtained similar values according to DPD data for the period 1982-2010. The mode of the occurrence frequency distribution for DPD lie for ratios equal to 0.15-0.2 and the median and mean values are 0.18 ± 0.06 and 0.22 ± 0.02, respectively. Therefore, those values are compatible within the error bars both for the median and mean. Instead, a small difference can be seen in the shape of the distributions where DPD distribution presents higher frequency values for low U/P. We have applied the Student's t-test with a 95 % confidence interval in order to study the similarity of the distributions. We found a statistical difference between both distributions. However, if we discard the U/P values lower than 0.1 in both distributions and consider a 99.9 % confidence interval, the obtained *t*-value (3.94) becomes close to the critical *t*-value (3.36). This implies that the difference between both distributions can be explained by the natural observational bias towards larger sunspots observed in the MM observations. There are other factors that can contribute to the difference: the low number of statistics in the MM distribution, the differences in the instruments employed to observe, and the large number of observers with respect to the 13 observers considered for MM distribution.

Furthermore, we have showed the evolution of the U/P for sunspots crossing the solar disk. We can observe the foreshortening effect on the measurements of several cases. A clear example of this kind is the sunspot observed by G.D. Cassini in 1676 November. This sunspot was registered as it traversed the disk and the values of U/P increased sharply near the solar limbs. Moreover, we compared the values of U/P for the same sunspot obtained from records of different observers. In the first case, we obtained a discrepancy around 25% in the ratio value and, for the second example, we found differences between 40% and 55% for the ratio values of the three sunspots examined from the de la Hire and Cassini (son) records. Finally, in order to check our methodology, we compared the values for the U/P obtained for sunspots observed by G.D. Cassini in 1676 October and November, which are included in three different historical documentary sources. We clearly see the good agreement between the results obtained from the three sources and therefore we can conclude that our method works well.

The result obtained in this work according to the mode, median, and mean values are in agreement with the ratio obtained for other epochs (although we have found statistical differences with respect to the modern DPD distribution). We note that Vaquero *et al*. (2005) found a ratio equal to 0.255 according to observations made in the Kew Observatory during the second part of the 19$^{th}$ century. This value is similar to the values found by other studies that use data from the 20th century (Nicholson, 1933; Waldmeier, 1939; Jensen *et al*., 1955; Zwaan, 1978; Brandt *et al*., 1990). Hoyt & Schatten (1997) pointed out that the rate of sunspot decay is proportional to the convective velocity indicating that higher convective velocities mean higher values for the U/P and faster sunspot decay rates. According to our result, the value of the ratio is similar to that for other epochs so the near absence of sunspots during the MM is not accompanied by changes in the U/P.

**Acknowledgment**

This work was partly funded by FEDER-Junta de Extremadura (Research Group Grant GR15137 and project IB16127) and from the Ministerio de Economía y Competitividad of the Spanish Government (AYA2014-57556-P and CGL2017-87917-P). The authors have benefited from the participation in the ISSI workshops. L. Lefèvre acknowledges support from BELSPO BRAIN VAL-U-SUN (BR/165/A3/VAL-U-SUN).

**Disclosure of Potential Conflicts of Interest**

The authors declare that they have no conflicts of interest.